# Microscopic model of the operation of the Single-chalcogenide X-point Memory

P. Fantini[1], A. Ghetti[1], E. Varesi[1], A. Pirovano[1], F. Pellizzer[1], D. Baratella[2], C. Ribaldone[2], S. Caravati[2], D. Campi[2], M. Bernasconi[2] and R. Bez[1]

[1]Micron Technology Inc., Vimercate, Italy, email: pfantini@micron.com

[2]Department of Materials Science, University of Milano-Bicocca, Milan, Italy

## Abstract

Ovonic threshold switching is the key process for several applications of chalcogenide alloys including phase change memories and selector elements in cross-points arrays. Very recently, it has been shown that the threshold switching voltage $V_T$ depends on the polarity of the applied field. This feature has been already exploited in the realization of the Single Chalcogenide X-point Memory (SXM) in which a single film of a chalcogenide alloy can serve as both a memory and selector unit. In this work, we provide a microscopic understanding of the polarity-dependent $V_T$ by leveraging electrical and physical measurements, numerical simulations based on technology computer aided design (TCAD) and electronic structure calculations based on density functional theory (DFT). We developed a Graded Band Gap (GBG) model in which an inhomogeneous distribution of localized electronic states in the gap is established by the opposite effect of a strong electric field at the cathode and a high density of electrons in the conduction band at the anode. The model is suitable to reproduce several features of the programming window, including its dependence on temperature, thickness and composition of the chalcogenide alloy. The microscopic understanding that we gained on the SXM operation lays the foundation for important improvements in the memory design and in the selection of better performing alloys for applications in enabling memory technologies.

## Introduction

Chalcogenide amorphous alloys are known to undergo a reversible electronic transition between a high-resistive state (OFF state) and a conductive state (ON state) upon application of a voltage exceeding a threshold value. This process is called ovonic threshold switching (OTS) after Ovshinsky who discovered this effect in the late 1960's [1]. Since early 2000, this feature is exploited in phase change memory (PCM) devices made of telluride alloys that also undergo a fast and reversible transformation between the amorphous and crystalline phases via Joule heating above the switching event [2, 3]. PCMs are being exploited in several enabling technologies such as storage class memories [4], embedded memories for microcontrollers [5], and devices for neuromorphic and in-memory computing [6].
More recently, OTS has attracted a widespread interest as a potential technology for selector devices [7, 8] in combination with both PCM [9,10] and resistive memories [11]. In the selector device, the chalcogenide alloy undergoes the OTS, but it does not crystallize. More recently, it was found that OTS selectors can also feature a threshold voltage ($V_T$) that depends on the polarity of the programming pulses [12, 13]. It was shown that $V_T$ differs if the reading pulse has the same or opposite polarity with respect to the programming pulse [12, 13, 14]. This phenomenon paved the way for the realization of memory devices where just a confined thin film of a chalcogenide glass plays the roles of both the selector and the information storage element. Memory cells exploiting this principle, called selector only memory (SOM), self-selecting memory (SSM), or Single-chalcogenide X-point Memory (SXM) have been presented by leading memory manufacturers [15, 16, 17]. A compact 2-terminals SXM device with physical size of less than 20 nm soon appeared as a prototypical self-selecting memory element that can be integrated in a 3D multi-tier NAND-like architecture [16]. The characteristics of SXM fit well with the requirements for storage class memories, as they combine the cost structure of a vertical 3D NAND with superior latency, endurance and bandwidth [16]. A sketch of the operational principle of SXM is shown in Fig. 1.

Despite the rapid technological exploitation of SXM, the physical mechanism behind its operational principle, i.e. the polarity dependent OTS, is not fully understood yet. It was initially proposed that this effect could arise from atomic segregation in the presence of strong electric fields and high currents in the ON-state with an accumulation of cations at the cathode and of anions at the anode [12, 15, 18]. The local change in composition would change either the Schottky barriers for carrier injection, or the local band gap, or the distribution of defect states within the band gap which controls the sub-threshold current and then $V_T$. However, evidence of atomic segregation was provided for GeSe only at very high operating current [18], while no segregation was found for switching at low current either in GeSe or in multi-components OTS materials like SiGeAsSe, as reported in a recent work by Ravsher et al. [19]. The same group recently proposed a new model in which the programming pulse generates in-gap states with an intrinsic anisotropy that respond differently to electric fields of opposite polarity [20]. More insight about the nature of these anisotropic structural defects is, however, not clear yet. Moreover, this model is rooted on a particular interpretation of the OTS mechanism itself which was based on the delocalization/percolation of some localized states in the band gap induced by the electric field at the threshold voltage [21, 22]. Another model has been proposed recently for the polarity dependent OTS [23] which is also based on the assumption that OTS is driven by a localized-delocalized transition of some trap in-gap states, although a microscopic identification of these key trap states is not available yet [23]. Most recently, Sung et al. [24] proposed a new model where a non-uniform distribution of hole and electron trap defects are formed by the strong electric field in the forming process of GeAsSe (GAS) alloys. The hole (electron) traps have been associated with Se-Se (cation-cation) homopolar bonds, by combined deep level transient spectroscopy measurements and electronic structure calculations based on density functional theory (DFT) [24].

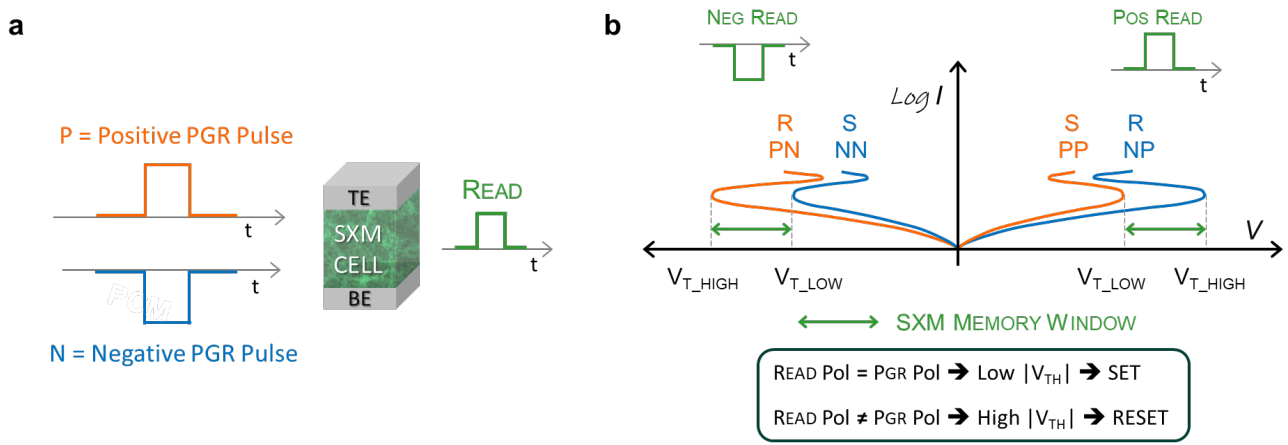

**Figure 1 | Operation of the SXM cell. a** Sketch of the positive (P) and negative (N) programming (PGR) pulses with a positive (P) polarity readout pulse. **b** I-V characteristics and $V_T$s illustrating the different programming and readout polarities. $V_T$ is higher when program and read occur in opposite polarity (NP and PN cases), while $V_T$ is lower when program and read have concordant polarity (PP and NN). The different $V_T$ provides the basis for the storage principle of the SXM device.

In the present work, we propose an alternative microscopic model for the operation of SXM based on experimental electrical measurements supplemented by DFT calculations and by numerical simulations based on Technology Computer Aided Design (TCAD). The new model, that we name graded band gap (GBG) model, is consistent with the most recent interpretation of the OTS mechanism itself which is based on a bipolar impact ionization avalanche mechanism [25]. In the presence of a different mobility of electrons and holes, this latter model for OTS is suitable, among other features, to account quantitatively for the observed *S*-type negative differential resistance (NDR) [25]. Once implemented within a TCAD simulation tool, the GBG model can match all the relevant experimental characteristics of the SXM window as a function of the thickness of the active material, of temperature, and of the alloy stoichiometry. A preliminary account of these results was reported in the proceeding of the 2024 International Electronic Device Meeting [26]. The model was developed after providing additional experimental data disproving the role of atomic segregation in the SXM operation, on which we report in the opening of the following section.

# Results

**Experimental data disproving the segregation model**

The SXM operational mechanism was originally interpreted as due to the atomic migration promoting the accumulation of anions (chalcogens, ch) and the depletion of cations close to the anode, and vice versa at the cathode. To mimic this compositional gradient, we fabricated a bi-layered chalcogenide glass device, with ch-rich and ch-poor regions, by using a low thermal budget process to inhibit layer intermixing (Fig. 2a). Ellipsometry measurements on each single layer confirmed the increase of the optical bandgap with chalcogen atomic concentration (SI Table S1). The *I-V* characteristics of such stacked capacitors have been measured under both polarities as shown in Fig. 2b. For the polarity in which the ch-poor layer is at the anode, corresponding to the readout of the RESET state, i.e. opposite polarity with respect to the programming pulse, the *I-V* curve shows a higher conductivity than in the other polarity (Fig. 2b). This result is opposite with respect to the behavior of a segregated-programmed cell where the RESET state shows lower sub-threshold leakage, associated with a higher $V_T$. Fig. 2c shows the activation energy for conduction ($E_A$) as a function of voltage, extracted from the *I-V* curves in Fig. 2b at increasing temperatures between 25 °C and 85 °C. The measured $E_A$ is higher for the SET state than for the RESET state, in agreement with the higher bandgap associated to the ch-rich region and lower band gap associated to the ch-poor region as reported in the inset of Fig. 2c. This result is opposite to the observed field-induced programming effect.

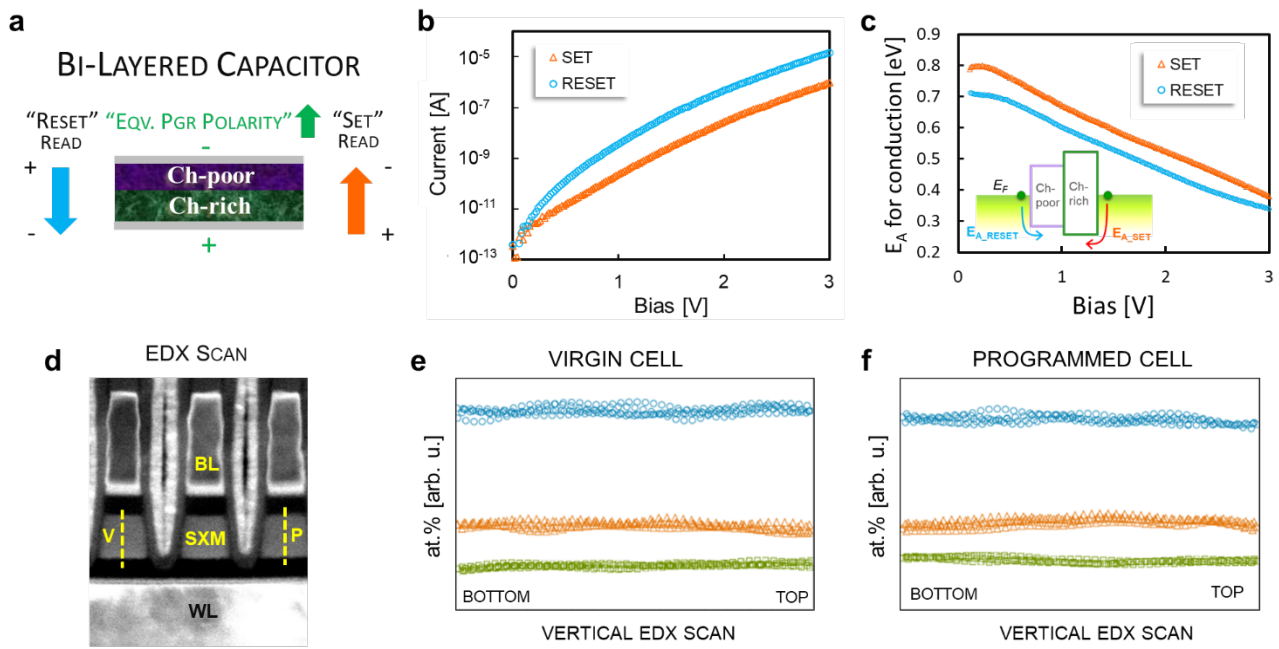

**Figure 2 | Experimental observations invalidating the segregation model. a** Bi-layered chalcogenide glass arranged in a capacitor structure to mimic a segregated cell. **b** Polarity dependent I-V curves from the bi-layered capacitor shown in panel **a**. **c** Activation energy for conduction ($E_A$) extracted from the polarity dependent I-V in **b** measured at different temperatures. **d** Transmission electron microscopy (TEM) image of a portion of the SXM array fabricated in 20 nm half-pitch technology including a virgin and a programmed cell. **e** Energy Dispersive X-ray (EDX) scan along the direction of the electric field and current for a virgin cell as in **d**. **f** The same of panel **e** for a polarity programmed cell. No segregation is detected for this specific stoichiometry.

To better assess the possible element segregation after field-induced switching, we conducted Energy Dispersive X-ray (EDX) on SXM devices [16] with different composition (Fig. 2d). For a particular cell composition, we found a wide SXM window, however no segregation could be obtained by the EDX scan, as reported in Fig. 2e-f. In summary, two independent experimental observations invalidate atomic segregation as the mechanism behind the SXM window proposed in Refs. [15, 18].

**The Graded Band Gap model**
We here discuss our new model for SXM which is based on the presence of a graded band gap (GBG). The formation of this feature in the cell is related to a different response of defect states in the gap to electric fields and to injected currents. The GBG model relies on the recent explanation of the OTS mechanism by a bipolar carrier multiplication model which is sketched in Fig. 3a [25]. The model is capable of reproducing the *S*-type NDR current-voltage characteristic by leveraging on three key ingredients, namely (i) the Poole-Frenkel conduction by majority holes at low electric field, (ii) the bipolar avalanche multiplication that triggers the switching regime, and (iii) the large secondary electron mobility compared to the hole mobility leading to the ON-state. The Poole-Frenkel conduction mechanism for holes requires the presence of localized in-gap states close to the mobility edge at the valence band. The band diagram in Fig. 3a of the ON-state switching highlights two anode-cathode asymmetries that occur during the electrical switching of the cell and that are reproduced by our model, i.e. the presence of high density of electrons at the anode and of a high-electric field at the cathode. This asymmetry is the pivotal element for our explanation of the SXM operation. In fact, as we will see in the next sections, atomistic simulations show that these different features at the two electrodes have a different impact on the in-gap localized electronic states. Namely, the injection of electrons in localized in-gap states close to the conduction band triggers an irreversible local structural transformation that brings the energy of these localized states closer to the valence band edge. This leads to a decrease in the activation energy for hole injection and then to a higher sub-threshold conduction. On the contrary, the presence of a strong electric field causes an irreversible removal of the localized states in the energy band gap that leads to an increase in the activation energy for hole injection and then to a larger effective band gap for conduction. The removal of defect states in the gap by an electric field is also consistent with the acceleration of the drift (increase) in the electrical resistance once the system is stressed by an electric field below threshold. The field-accelerated drift is well documented in literature for the $Ge_2Sb_2Te_5$ phase change alloy [27], and a similar phenomenology is expected to hold for OTS selector alloys as well.
Therefore, the two combined effects of high density of injected electrons at the anode and high electric field at the cathode give rise to a graded band gap with a different activation energy for hole injection at the electrodes. The direction of the band gap gradient is defined by the programming step as sketched in Fig. 3b and it is responsible for the polarity dependent OTS. In fact, when the read pulse has the same polarity of the programming pulse, holes are still injected at the anode where the activation energy is lower and then $V_T$ is lower as well. On the contrary, when the read pulse has the opposite polarity, holes are injected at the cathodic side of the programming step where the activation energy is higher which leads to a higher $V_T$. Any switching event refreshes the system by changing the graded band gap according to the programming polarity. This is possible because in the ON-state the system is brought above the glass transition temperature where atomic mobility is sufficiently high to restore a uniform distribution of defects before the combined effects of the high electric field and of electrons injection at the two electrodes freeze the out-of-equilibrium asymmetry created by the switching event, where different structure relaxation and bang gap recovery take place at the two opposite sides of the chalcogenide glass.
In the next section, we will show that the GBG model is able to reproduce the main features of the OTS process in SXM including the dependence of the SXM window on temperature, composition and chalcogenide thickness. Atomistic support for the GBG model from DFT simulations will be presented in the subsequent section.

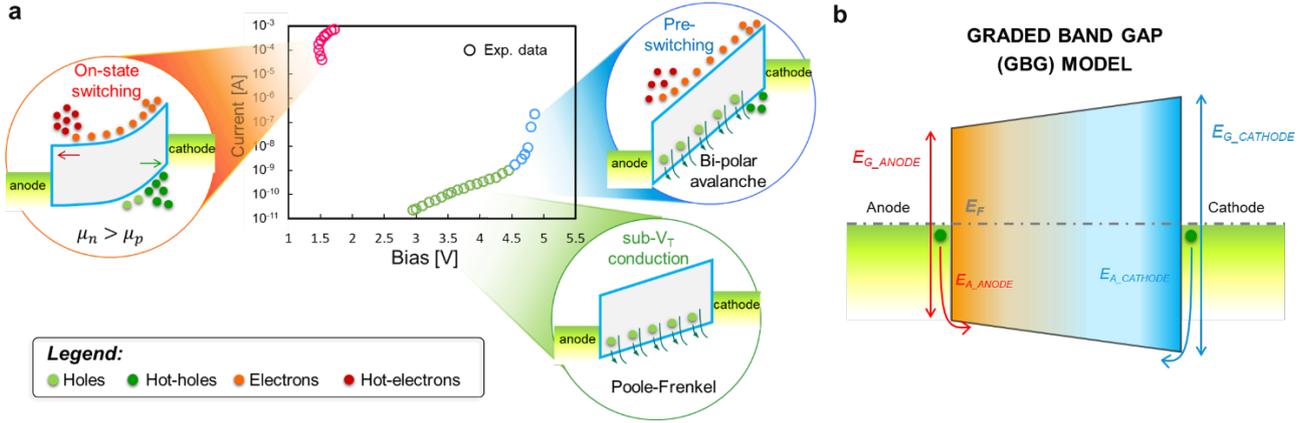

**Figure 3 | Threshold switching model and graded band gap (GBG) model. a** Simulated *S*-shaped current-voltage (*I-V*) characteristic by means of the bipolar avalanche model. The schematic band diagrams surrounding the plot describe qualitatively the different conditions: from the low field pre-switching regime up to the bipolar multiplication avalanche that gives rise to the NDR region and the ON-state (adapted from Ref. [25]). **b** Graded Band Gap (GBG) model responsible for SXM window with the two different activation energies ($E_A$) for hole injection at the anode and cathode.

### TCAD simulations of the switching process in the graded band gap model

The GBG model has been implemented in a TCAD simulation tool to simulate the SXM operation. The sub-threshold conduction has been described by means of a Tunneling-Assisted Poole-Frenkel (TAPF) mechanism in which carriers with a density of states *f(E)* are injected into the chalcogenide glass by thermally assisted tunnelling process from the electrodes, while the bulk mobility is controlled by trap-to-trap Poole-Frenkel hopping, potentially involving intermediate transitions into conduction/valence states outside the mobility gap (Fig. 4a) [28]. Such a conduction model also agrees with the polarity dependence of the *I-V* curves of the bi-layered capacitor reported in Fig. 2. In fact, the activation energy for conduction $E_A$ corresponds to the value of the Schottky barrier height of the electrode/chalcogenide junction, which is lower at the ch-poor interface and higher at the high $E_G$ side (ch-rich interface, because of the increase in the band gap with the fraction of chalcogen). Moreover, the bipolar avalanche OTS model has been adapted by considering tilted bands with a band gap gradient ($\nabla E_G$). The current *I* depends on the distance *u* from the injection electrode as $I = I_o\, e^{\alpha u}$, where $\alpha$ is the impact ionization (II) coefficient depending in turn on both $\nabla E_G$ (Fig. 4a) and the electric field *F*, as given by

$$\alpha = A_{e/h}\exp\left(-\frac{E_G(x)}{q\lambda\left(F \pm \nabla E_G\, \xi_{e/h}\right)}\right) \quad (1)$$

Here $A_{e/h}$ is a pre-exponential factor for electrons or holes, *q* is the electron charge and λ is the mean free path for II [29]. In particular, the opposite sign of the band gap gradient experienced by electrons with the flipped readout polarity provides the correct SET/RESET $V_T$ dependence (Fig. 4 b, c). Overall, our model coherently describes the experimental correlation between leakage current and $V_T$, i.e. a higher sub-threshold current and lower $V_T$ in the SET than in the RESET state. The model contains just a few parameters, $A_{e/h}$, $\lambda$ and $\xi_{e/h}$, that have been tuned within a range of physically meaningful values to fit the experimental trends of SXM cells. Fig. 5a reports the comparison between the simulated and experimental *I-V* characteristics for SET and RESET by varying temperature. The change of $V_T$ with temperature in the SET and RESET states is well described by the calibrated temperature dependence of the impact ionization pre-factor $A_{e/h}$ (Eq. (1)) in the two states. The model can also reproduce the linear increase of the SXM window with the thickness of the chalcogenide glass as shown in Fig. 5b. This variation can be understood by considering that the critical value of impact ionization coefficient $\alpha$ dictating the onset of switching increases by decreasing the thickness [25] and that the GBG model yields a different slope of $\alpha$ with the inverse of electric field in the SET and RESET states as shown in Fig. 5c. These two combined effects lead to a lowering of the difference in the threshold electric field for the SET to RESET by decreasing thickness as illustrated in Fig. 5c which ultimately reduces the memory window. Finally, Fig. 5d shows the dependence of polarity window on the alloy ionicity calculated as proposed in ref. [30]. In this context, the higher is the alloy ionicity the larger is the effect of the electric field on the removal of the wrong bonds (see next section) responsible of defect states in the gap which is at the basis of the generation of the SXM window. As we will elaborate further in the next section, defects formation and annihilation modulate the graded band gap that ultimately controls the SXM window. By introducing a linear relation between the gradient of the band gap ($\nabla E_G$) and the alloy ionicity as sketched in the inset of Fig. 5d, TCAD simulations well reproduce the experimental behavior (Fig. 5d).

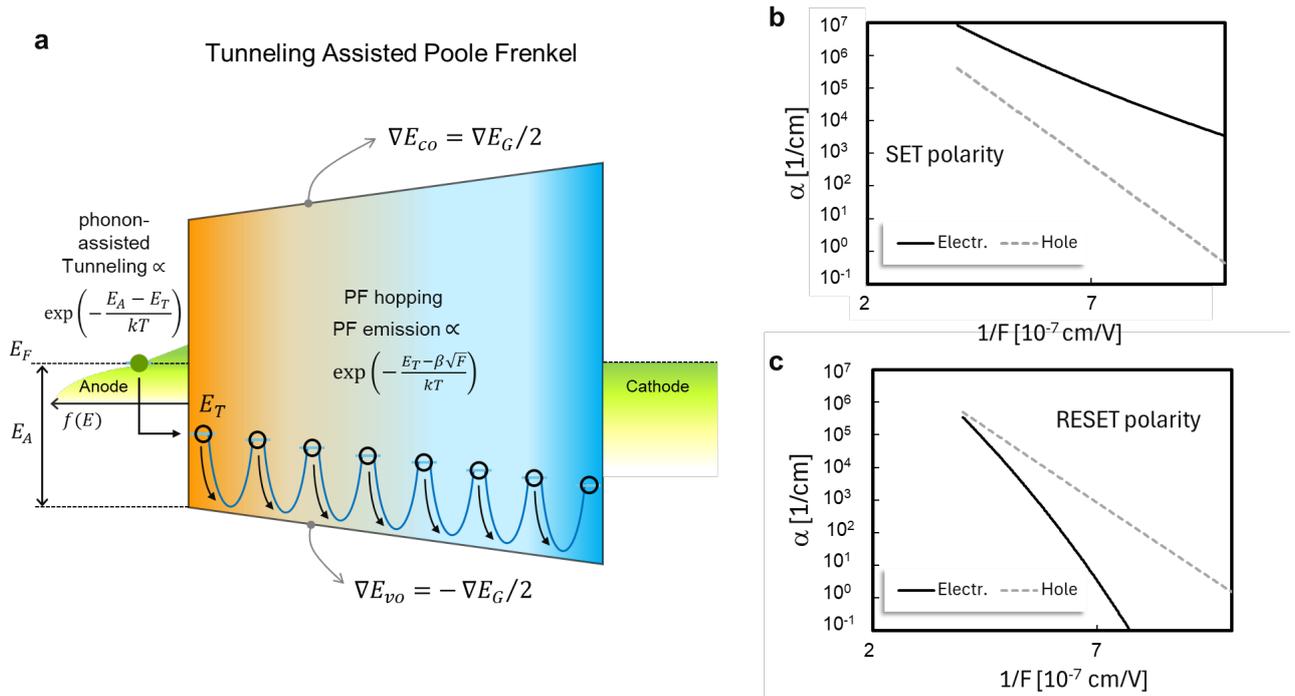

**Figure 4 | Subthreshold conduction mechanism and impact ionization coefficient. a** Energy-band diagram with a graded band gap giving rise to the gradient of the conduction and valence band edges $\sim\nabla E_G/2$. A sketch is given of the Tunneling-Assisted Poole-Frenkel (TAPF) model implemented in our TCAD simulator to describe the sub-threshold current. Estimated impact ionization coefficient $\alpha$ as a function of *1/F* for electrons and holes for **b** SET and **c** RESET. $\nabla E_G$ attenuates the $\alpha$ coefficient for electrons that slowdowns the bipolar avalanche onset for the RESET leading to higher $V_T$.

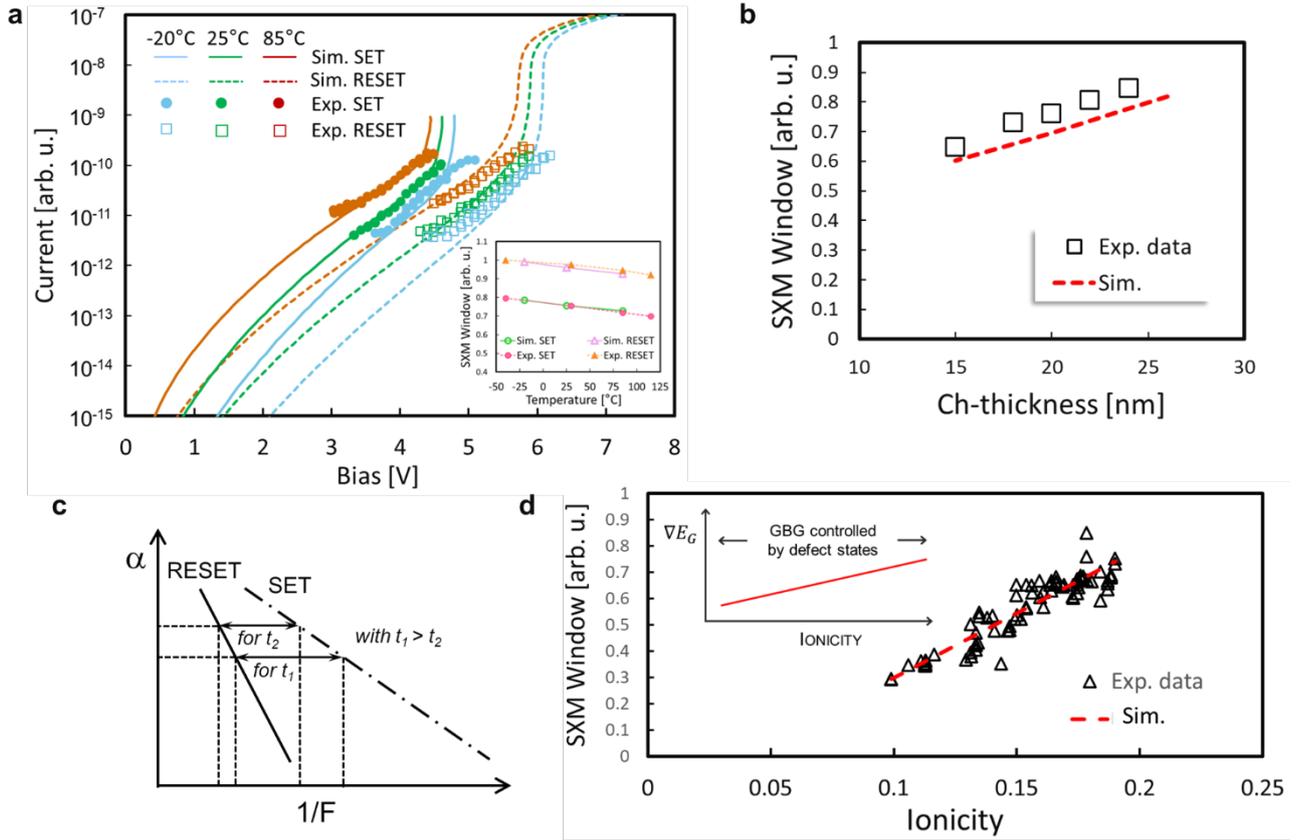

**Figure 5 | TCAD simulations vs experimental trends. a** Simulated and experimental I-V characteristics for SET and RESET by varying temperature. The simulated and experimental dependence on temperature of $V_T$ for SET and RESET are shown in the inset. **b** SXM window as a function of the thickness of the amorphous chalcogenide film: experimental data and TCAD simulation results are compared. **c** Impact ionization coefficients for SET and RESET as a function of $1/F$; the critical value of $\alpha$ for OTS is shown for two different thicknesses of the chalcogenide film $t_1 > t_2$. **d** Experimental SXM window vs. alloy ionicity calculated as in ref. [30] and TCAD-simulated trend by introducing in the model the linear relation in $\nabla E_G$ sketched in the inset.

## Atomistic simulations of localized states in the band gap

To support the GBG model described above, we performed DFT molecular dynamics (MD) simulations and analyzed the evolution of the defect states in the gap either under electron injections or application of a strong electric field. As a prototypical OTS material, we considered the ternary GeAsSe alloy which is among the most promising materials for selector applications [8, 31]. In particular, we chose the composition $Ge_{25}As_{30}Se_{45}$ (GAS234), i.e. an off-stoichiometric alloy that lies on the pseudobinary line between Ge and $As_2Se_3$, namely $Ge_{25}(As_2Se_3)_{15}$. This choice resulted from an extensive DFT analysis of the nature of in-gap states and their dependence on composition in several GeAsSe alloys that will be discussed in a separate publication. As we will see below, at this composition the in-gap states are mainly localized on Ge-As and As-As bonds. A comparison with other DFT works on the most studied $GeSe_x$ alloys will be discussed later on in this section. Four amorphous models of GAS234 (models M1-M4) were generated in a 300-atom cubic box by quenching from the melt to 300 K within DFT-MD, as described in the Method section. Information on the structural properties of the amorphous models (pair correlation functions, distribution of coordination numbers, partial coordination numbers, percentage of the different types of bonds) are given in Figs. S1-S2 and Tables S2-S3 in the Supplementary Information. Se atoms are mostly 2-coordinated with a significant fraction of 3-coordinated atoms; As atoms are 3-coordinated in an overwhelming fraction; Ge atoms are 3- or 4-coordinated with very few atoms 5-coordinated. The geometry of the different local environment is shown in Fig. S3 in the Supplementary Information with the type of bonding highlighted by the Wannier functions [32]. 3-coordinated atoms of all species are in a pyramidal geometry, while the large majority of 4-coordinated Ge atoms corresponds to a tetrahedral bonding geometry with sp$^3$ hybridization (see Fig. S3 in the Supplementary Information). The 5-coordinated and a minority of 4-coordinated Ge atoms are in a defective octahedral geometry (octahedral bonding angles, but coordination lower than six). The fraction of Ge atoms in a tetrahedral geometry and the fraction of different types of tetrahedra (isolated, corner- and edge-sharing) are given in Table S4 in the Supplementary Information. As shown in Table S3, the majority of bonds are Ge-Se (49.5 %) and As-Se (27 %), but with a non-negligible fraction of As-As (15.5 %) and Ge-As (5.7 %) bonds, while the fraction of Ge-Ge homopolar bonds (2.5 %) is very low. Homopolar bonds and Ge-As bonds are not present in the crystalline form of ternary and binary compounds in the Ge-As-Se phase diagram, and they are therefore often referred to as wrong bonds. A snapshot of model M1 of amorphous GAS234 (a-GAS234) is shown in Fig. 6a. We remark that both GeSe and GeSe$_2$ exist as stoichiometric crystalline compounds. GeSe crystallizes in an orthorhombic phase (space group Pnma) [33]. The crystal is made of GeSe bilayers in which atoms are 3-fold coordinated in a pyramidal configuration. Interbilayer bonds are mostly due to van der Waals interactions. The formation energy is 60 kJ/mole and the electronic gap is 1.09 eV [33]. GeSe$_2$ crystallizes instead in a monoclinic phase (space group P2$_1$/c) [34] made of chains of corner- and edge-sharing tetrahedra. The formation energy of 115 kJ/mole and the electronic gap of 2.18 eV [35] for GeSe$_2$ are both larger than in GeSe. For GeSe$_x$ alloys the tetrahedra are thus lower in energy than the pyramidal configurations. The electronic density of states (DOS) for model M1 is superimposed to the inverse participation ratio (IPR) of individual Kohn-Sham (KS) orbitals in Fig. 6b (see Methods). The IPR gives a measure of the localization of the KS orbitals as it takes values ranging from 1/N for a completely delocalized electron, where N is the number of atomic-like orbitals in the basis set of the whole supercell, to one for an electron completely localized on a single atomic-like orbital. The state with the highest IPR close to the conduction band edge is localized over a chain of wrong As-As, and As-Ge bonds (As-Se-As-As-As-Ge chain, see inset of Fig. S4 in the Supplementary Information). Similar features are obtained in the other three independent models of a-GAS234 (models M2-M4) in which the in-gap states with higher IPR are all mostly localized on wrong bonds.

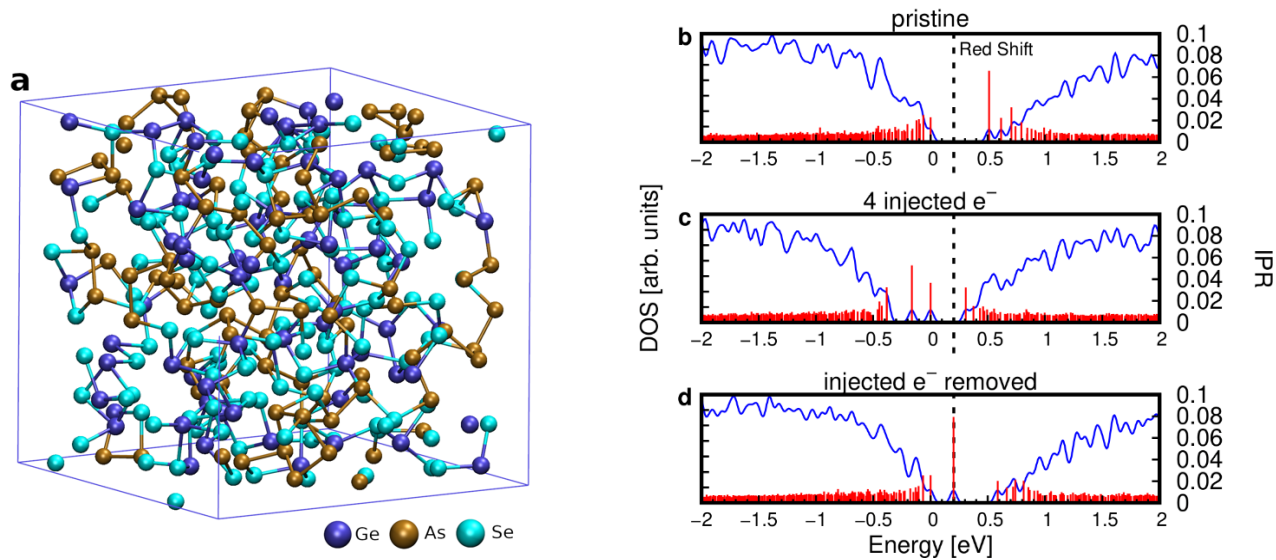

**Figure 6 | Snapshot of the atomistic model of amorphous Ge$_{25}$As$_{30}$Se$_{45}$ and its electronic density of states with and without injected electrons. a** Snapshot of the atomistic model of a-GAS234. **b** Electronic density of states (DOS) and inverse participation ratio (IPR, red spikes) of model M1 of a-GAS234. The zero of energy is the highest occupied state (top of the valence band or highest occupied molecular orbital, HOMO). The DOS is computed from KS energies at the supercell Γ-point with a 27 meV Gaussian broadening. **c** DOS and IPR with 4 additional electrons and **d** after removal of the added electrons. The system is relaxed after addition or removal of electrons in panels **c** and **d**. The red shift of the in-gap localized state is the signature of the electron-induced irreversible transformation.

We have then injected additional electrons to our model M1 (compensated by a uniform positive background) aiming at mimicking the condition at the anode during the switching operation, with a high concentration of injected electrons (see section above on the GBG model). The injection of one additional electron in our cell corresponds to an electron density of about $1.2\ 10^{20}/cm^3$ which it is not too far from the estimate of the density of injected electrons above the threshold in the ON-state of about $6\ 10^{19}/cm^3$ given by the phenomenological model of Ref. [36]. We thus optimized the geometry of the amorphous model by adding 2-6 electrons at 0 K. Then, we removed the additional electrons, and we relaxed the system again to see if the transformation induced by the added electrons could survive once the electrons are removed. In Fig. 6, we compare the electronic DOS of the original model (panel b) with the DOS of the system after relaxation with 4 additional electrons (panel **c**) and after removal of the added electrons and further relaxation (panel d). The two empty states close to the conduction band at about 0.5-0.6 eV in Fig. 6b shift close to the valence band and below the zero of energy (highest occupied molecular orbital, HOMO) once four electrons are added and the system is relaxed (Fig. 6c). A self-trapped polaron forms for both states. The energy of these KS states shifts even closer to the valence band when the electrons are removed at clamped ions, because the energy of an individual KS state depends on its occupation, shifting towards higher energy the larger is the number of electrons that occupy the state at clamped ions. Once the four electrons are removed and the system relaxed (Fig. 6d), the initial DOS is not recovered which means that the additional electrons have induced irreversible structural transformations that survive once the added electrons are removed. Notice that the lowest unoccupied molecular orbital state (LUMO, i.e. the first empty state) shifts from the edge of the conduction band in the original system to a position close to the valence band after addition and then removal of four electrons. This state keeps its character during the transformation, i.e. a localization on a chain of wrong As-As and As-Ge bonds, as already mentioned above. The transformation leads to local structural relaxations with an energy gain with respect to the original model. Similar results are obtained by adding 2 or 6 electrons. A similar behavior is observed by charging a second independent model of a-GAS234 (model M2). In the third model instead (model M3), the transformation induced by up to six additional electrons is similar but mostly reversible, the DOS of the original model is in fact recovered after removal of the added electrons and relaxation. We must, however, consider that we used a protocol of geometry relaxation at zero temperature that can just find local minima with no activation barrier to overcome. The real system at finite temperature is expected to undergo a transformation lasting up to a few ns in the presence of an electronic current. This time scale is sufficiently long for the system to overcome activation barriers larger than the thermal energy with a sizeable probability. Moreover, we remark that both the initial configuration after quenching and the final configuration after injection and removal of additional electrons are off-equilibrium states. In an ideal glass generated by a quenching/annealing sufficiently slow, we do not expect to see localized states in the gap at all. Finally, we mention that in the fourth model (M4) the shift to lower energy of the LUMO state is reversible when it is filled by two electrons. However, when the system was relaxed after addition and removal of six electrons, this LUMO state disappeared from the gap. This state was initially localized of an under-coordinated (two-fold) As atom within a chain of wrong bonds. The initial DOS was then mostly recovered but for the disappearance of this state due to an irreversible structural transformation occurring only at high electron charging (see Methods for further details on this model).

In summary, we have seen examples in which the addition of electrons in localized states in the gap leads to the formation of self-trapped polarons with structural relaxations that survive upon removal of the additional electrons and that bring these empty states closer in energy to the valence band. Therefore, this outcome shows that the electron accumulation at the anode (see Fig. 3a) leading to the occupation of initially empty states can result in an increase of the subthreshold current due to the shift of empty localized states closer to the valence band. This feature supports the first assumption of GBG model on the behavior at the anode. We now briefly comment on previous works in literature on the mid-gap states in amorphous models of $GeSe_x$ alloys generated by quenching from the melt in DFT-MD simulations. We remark that the nature of the mid-gap states is expected to depend on composition (see Ref. [8] for a review). Valence-alternation pairs or other Se-related states have been reported in Se-rich $GeSe_{1+x}$ alloys [21, 37], while in $Ge_{50}Se_{50}$ and Ge-rich $Ge_{1+x}Se$ alloys the in-gap states, either deep or at the band edges, are related to Ge-Ge homopolar bonds [21]. The presence of a single Ge-Ge bond or a chain of Ge-Ge bonds is obviously not sufficient to give rise to a localized state. These structural features should in fact appear in combination with suitable medium range order (MRO) and under-/over-coordination as shown by the analysis of several models in Refs. [38, 39]. Since Ge-Ge bonds are weaker than Ge-Se bonds, we can conceive that homopolar bonds are more prone to accumulate local stress due to MRO and therefore to assume less favorable bond length, angle, or local environment giving rise to localized states in the gap. The formation of a polaron-like state in empty localized states after charging with an additional electron was reported in Ref. [38] for GeSe. In Ref. [40], the charging of electron traps was instead induced by a very high density of electron-hole pairs, involving also the band edges, that leads to a closing of the band gap itself. This sort of metallization process was associated to the increase of the fraction of axial bonds (with bond angles of about 180º) and it was proposed as the mechanism of OTS process itself [41]. Similar behavior was reported for GeSbSe alloys in which the increase in the fraction of axial bonds upon electronic excitation was mostly related to Sb [41]. Similar reduction in the band gap was observed in Ref. [42] by adding additional electrons in a model of amorphous $Ge_2Se_3$. Our results on the redshift in energy of the localized in-gap states once they are charged by electrons due to irreversible structural transformation is overall consistent with the previous accounts in literature on the other selenide alloys summarized above.

We have then considered the effect of an external uniform electric field on localized states in the gap. A uniform electric field is introduced within the Berry's phase approach by keeping periodic boundary conditions (see Methods) [43] in models M1-M4. We first relaxed the system in the presence of an electric field of increasing strength (0.1, 0.25 and 0.5 V/nm). Then, we removed the field, and we relaxed again the system to investigate eventual irreversible transformations induced by the electric field. We observed a different behavior in the four models. In model M4, the largest electric field (0.5 V/nm) was able to remove irreversibly a mid-gap state (LUMO) localized on an under-coordinated (two-fold) As atom within a chain of wrong bonds (see above). The electronic DOS and IPR of model M4 before and after application of the electric field (and relaxation as described above) are compared in Fig. 7a-b. Similarly, a 0.5 V/nm electric field was suitable to remove irreversibly some mid-gap states in model M3 (see Fig. S5a in the Supplementary Information). In model M1 we observed a non-monotonic behavior of the in-gap states with an overall change which is rather marginal at the highest field (see Fig. S5b in the Supplementary Information). In model M2, the behavior is more complex with some states moving up and other moving down in energy (see Fig. S5c in the Supplementary Information). Overall, we do observe in two models out of four a removal of in-gap states induced by the electric field which supports the second assumption of the GBG model on the behavior at the cathode.

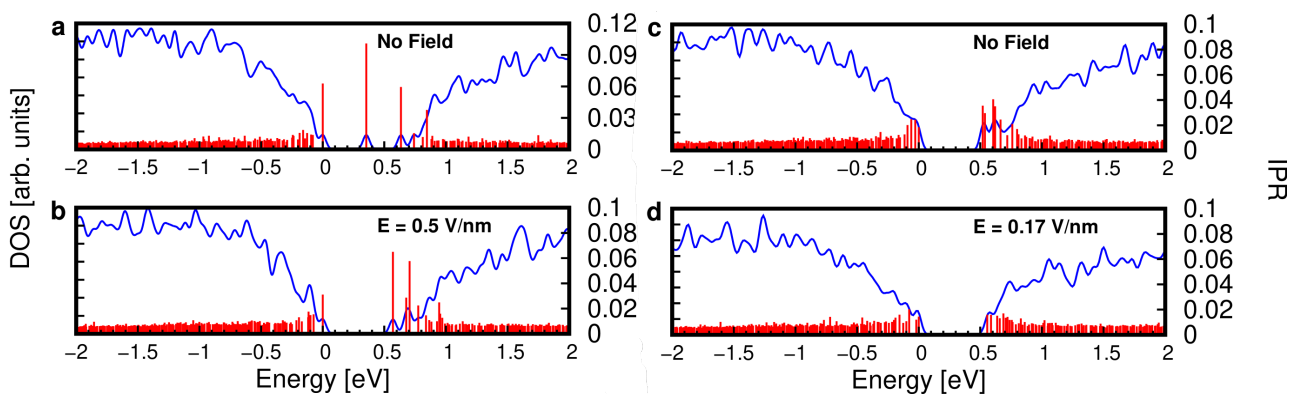

**Figure 7 | Effect of the electric field on the electronic density of states of a-GAS234.** Electron density of states (DOS) and Inverse Participation Ratio (IPR, red spikes) of **a** model M4 of a-GAS234 before and **b** after the application and then removal of a 0.5 V/nm electric field. The system was relaxed in the presence of the electric field and relaxed further after the electric field was removed. **c** DOS and IPR for a single snapshot of a model at 300 K generated by slowly quenching from 500 K to 300 K in the lack or **d** in the presence of an electric field of 0.17 V/nm. The DOS are then computed *without* the electric field for all configurations.

Note that this analysis is based on geometry optimization in the presence of the electric field but at zero temperature. Thus, the structural relaxations leading to the irreversible shift of the in-gap states can take place only if the field removes the energy barrier for the process to occur. The same barrier might be overcome at lower field at finite temperature. To assess this effect, we perform additional simulations in which we switched on the electric field during the quenching from the supercooled liquid phase. The simulations in the presence of an electric field within the Berry's phase approach (see Methods) can be accomplished only in the presence of a sufficiently large band gap between delocalized states. In the lack of a band gap sufficiently large to prevent a Zener breakdown [44], self-consistency in the solution of KS equation cannot be achieved. Since the supercooled liquid is metallic while the amorphous phase is semiconducting, a metal-semiconductor transition occurs upon cooling. We checked that simulations with a lower electric field of 0.17 V/nm can be performed smoothly up to 500 K. Therefore, from the equilibrated system at 500 K, we started two different simulations at constant temperature about 35 ps long, the first with no electric field and a second with an electric field of 0.17 V/nm. Then, we quenched abruptly to 300 K different amorphous models equilibrated at 500 K with or without the electric field. We generated four independent models (Q1-Q4) obtained by quenching from snapshots of the trajectory at 500 K sufficiently separated in time, i.e. at 5, 13, 23, and 35 ps. The models were then equilibrated for about 15 ps at 300 K and finally optimized at zero temperature. The quenching to 300 K was performed for all models *without* the electric field to mimic a final relaxation at zero bias after the system was stressed by the electric field. The DOS were consequently computed *without* the electric field for all models as shown in Fig. S6 in the Supplementary Information. Overall, one can notice that the electric field has caused a reduction in the number of empty localized states in the gap. Only in model Q1 one finds the enhancement of the localization of a filled state close to the valence band. On the contrary, two empty localized states at about 0.5 eV above the HOMO in model Q1 and one empty localized state at about 0.4 eV in model Q2 disappear in the presence of the electric field. All these states are localized on Ge-Ge and Se-Se bonds. The single KS state at about 0.4-0.5 eV in model Q3 change character from being localized on a chain of Ge-As-As bonds without the field to a state mostly localized on an As-Se bond with the field. The localized states close to the conduction bands in model Q4 also disappear in the presence of the electric field and they are instead mostly localized on Ge-As bonds.

Despite the limited statistics, we can attempt to rationalize these outcomes as follows. In an ideal glass we expect to have a prevalence of the stronger As-Se and Ge-Se bonds. These bonds are highly polar due to the difference in electronegativity of the atoms involved (Pauling electronegativities are 1.8, 2.0 and 2.4 for Ge, As and Se atoms). Homopolar Ge-Ge, As-As and Se-Se bonds and less polar Ge-As bonds are weaker and structural transformations, also responsible for aging, are supposed to remove these wrong bonds leading to the formation of the more polar As-Se and Ge-Se bonds. We can thus speculate that the effect of the electric field is a further stabilization of the polar bonds which allows for a reduction of the activation barriers to be overcome for the removal of wrong bonds. The analysis of the DOS is thus consistent with the removal by the field of a few, possibly less stable, Ge-Ge, Se-Se, As-As and Ge-As bonds responsible for the localized states in the gap.

We remark that the differences in the models Q1 - Q4 with and without the electric field must be ascribed to differences already present in the supercooled liquid at 500 K because the models are generated by an abrupt quenching from 500 K to 300 K. To investigate further the structural relaxations that might intervene at lower temperatures, we have also generated two different models by quenching from 500 K to 300 K in four steps for an overall simulation time of 60 ps with or without the electric field. The model generated with the electric field is further equilibrated at 500 K for other 12 ps before the quenching. The electronic DOS of a single snapshot at 300 K of the two models are shown in Fig. 7c-d. The DOS are computed *without* the electric field. The states at the band edges are less localized that in the models generated by an abrupt quench to 300 K because the system had more time to relax. Still the electric field leads to an overall widening of the band gap in this latter model as well. We can conclude that an electric field favors the removal of less stable local configurations (less polar bonds) responsible for localized states in the gap. Once more, this feature supports the second assumption of the GBG model on the behavior at the cathode.

We now briefly comment on previous works on the DFT simulations of the effect of the electric field on the in-gap states in chalcogenide alloys. In [21] and references therein from the same group, it was shown that the application of an electric field of 0.87 V/nm at clamped ions within the Berry's phase approach leads to a very large Stark effect in GeSe with an almost closing band gap. This behavior would be consistent with the field-induced delocalization of tail states discussed for instance in Ref. [45]. In this respect, we remark that in the present article we have always considered the change in the energy of localized states due to the structural relaxations induced by the electric field and not the energy shift (Stark effect) due to the field itself at clamped ions. All the electronic DOS that we have shown in the present article have been computed without the electric field. We also mention that a band gap closure similar to that found for GeSe by Clima et al. [21], has been found in a more recent study of $GeSe_2$ [37], in which an electric field of increasing strength up to 1.5 V/nm was applied during an MD simulation. This behavior cannot be ascribed to a simple Stark effect that could fit the energy shift at low field but not at high field, where structural changes induced by the field are also important. This effect is, however, reversible as the mid-gap states closing the gap annihilate once the field is removed [37]. Moreover, a significant redshift of the localized states is observed for fields of and larger than 1 V/nm which is about five times larger than the typical threshold field in selectors. We must also remark once more that the nature of localized defect states is different in Se-rich $GeSe_2$ and GeSe. In this respect, we also mention that the application of an electric field of 0.25 V/nm in DFT models of $Ge_2Sb_2Te_5$, induces structural relaxations that lead to the removal of mid-gap states [46].

## Discussion

In summary, we have developed a microscopic model for the operation of the single chalcogenide X-point memory suitable to explain the dependence of the threshold voltage $V_T$ on the polarity of the applied electric field. The model is consistent with our most recent understanding of the mechanism of OTS which is based on a bipolar impact ionization avalanche with a different mobility of electrons and holes via a trap-to-trap conduction process, through states outside the mobility gap [25]. The present model for SXM operation was inferred from both experimental measurements and DFT electronic structure calculations. Electrical characterization of a cell made of two chalcogenide slabs with different compositions allowed us to exclude that the polarity-dependent $V_T$ was due to a de-mixing of the alloys as previously proposed in literature. We then proposed a Graded Band Gap (GBG) model in which an inhomogeneous distribution of trap states (localized electronic states in the gap) arises from the combined effect of an electric field at the cathode and a high density of electrons in the conduction band at the anode. DFT calculations support the GBG model as they show that injected electrons could lead to a redshift of the in-gap states which come closer to the valence band via a polaronic effect, while a strong electric field can remove or blueshift in-gap states. It was also shown that TCAD numerical simulations implementing the GBG model are suitable to reproduce several features of the SXM window (difference between SET and RESET $V_T$) such as the dependence on temperature and on the thickness and ionicity of the chalcogenide. The microscopic understanding of the SXM device provided by the GBG model will be extremely useful for the optimization and scaling of the memory cell and for future exploration of different alloy composition in the perspective to exploit this architecture for storage class memories and other advanced devices.

## Methods

**Experiments**
Capacitor structures with bi-layered chalcogenide reported in Fig. 2 have been fabricated by PVD RF sputtering deposition of all the materials in the Carbon/GAS/Carbon stack and then by patterning the capacitor structure through a standard litho-and-etch process, granting an overall reduced thermal budget to prevent process-induced elemental interdiffusion in between the two chalcogenide layers. Composition of each chalcogenide sublayer was measured through X-Ray fluorescence spectroscopy (XRF) and optical bandgap (BG) was extracted by Tauc plot from ellipsometric measurements on each single sublayer. Optical BG results consistently higher for ch-rich alloy compared to ch-poor alloy (Table S1 Supplementary information). Capacitor with size from 50 to 150 $\mu$m was fabricated with GAS bilayer consisting of two sublayer 12 nm thick each. Quasi-static I-V curves have been collected with a Keysight B1500A semiconductor device parameter analyzer by forcing a voltage on the top electrode and measuring the current entering the top electrode. Device testing temperature was controlled through a thermos-chuck and measurements have been performed at different temperatures in the range 25–85 °C. Corresponding conduction activation energy was extracted based on Poole-Frenkel model. Samples with Se-poor/Se-rich stack sequence pretty overlaps with data of reversed one, excluding significant impact of the deposition sequence. The maximum applied voltage has been limited to stay well below the threshold voltage of these devices. TEM lamella has been prepared with FIB (ThermoFisher Helios 5 DualBeam) on few cells of a large Xpoint SXM array device with 20 nm half pitch [16] in virgin and programmed states (both polarities). Cell EDX TEM analysis has been performed by means of JEOL JEM-ARM200F microscope equipped with ThermoFisher Noran microanalysis EDS detector; element profiles between the electrodes of four virgin and programmed cells are reported in Fig. 2e-f. All the electrical data reported in Fig. 5 have been measured on 20 nm SXM cells. Quasi-static I-V curves have been measured with the same experimental setup adopted for the capacitor structures in Fig. 2. For the $V_T$ and SXM window measurement, voltage programming and reading pulses have been applied with a Keysight B1500A semiconductor device parameter analyzer (equipped with a HV-SPGU high-voltage pulse generator unit). Applied pulses have been measured on a Tektronix DPO7254 5600 digital oscilloscope. At the same time the current flowing through the cell is measured on a second channel of the scope, thus allowing to detect the exact timing of the transition from the OFF-state to the ON-state and to quote the corresponding applied voltage (threshold voltage $V_T$). Since the OFF-state current just below threshold is around some hundreds of nA, the voltage drops caused by the current flowing on the external routing to connect the cell is in the range of few mV and it can be easily neglected. SXM window is measured by detecting the $V_T$ of a first programmed high-$V_T$ state and then reprogramming and reading the same cell in the low-$V_T$ state.

**TCAD simulations**
TCAD simulations were carried out with the commercial software Sentaurus Device by Synopsys [47]. It solves the classical Drift-Diffusion equations for semiconductor transport self-consistently with the Poisson equation for the potential. The TAPF mechanism in Fig. 4a was included in the simulation by instantiating traps in the chalcogenide region coupled with the contact regions via nonlocal tunneling and activating the standard PF model for carrier emission from traps. Implementation of the Graded Band Gap model (Fig. 4a) and, especially, of the band edge gradient effect on the impact ionization coefficient (Eq. 1) was not trivial as required some development to extend the capability of the simulation tool. I-V simulations were done in stationary conditions. We used current boundary conditions to allow simulating of the snap-back. However, sometime simulations failed to converge once in the snap-back region because of the difficult numerical problem in this condition (Fig. 5a). Nevertheless, the threshold point was always visible allowing to correctly compute the SXM memory window.

**DFT simulations**
We performed DFT-MD simulations using the CP2k suite of programs [48] with norm-conserving pseudopotentials with four, five, and six valence electrons for Ge, As, and Se [49]. The Kohn-Sham orbitals were expanded in the basis set of triple-zeta-valence plus polarization Gaussian-type orbitals (GTO), while the charge density was expanded in a planewave basis set with a cutoff of 240 Ry to efficiently solve the Poisson equation within the Quickstep scheme [48]. MD simulations in supercells with periodic boundary conditions were performed by using the predictor-corrector scheme of Refs. [50] for the self-consistent solution of KS equation at each MD step. In the spirit of the Car-Parrinello (CP) approach [51], the wave functions are not self-consistently optimized during the dynamics. However, in contrast to CP, large integration time steps can be used in the simulation. This scheme leads to a slightly dissipative dynamics of the type $-\gamma \dot{R}_I$ where $R_I$ are the ionic coordinates. In Refs. [50] it is shown how to compensate for this dissipation and obtain a correct canonical sampling by applying a Langevin thermostat. The same scheme was applied in our previous works on the amorphous phase of telluride alloys [52]. MD simulations have been performed at constant volume (fixed simulation cell) and constant temperature enforced by a Langevin thermostat as mentioned above. Brillouin Zone integration was restricted to the supercell Γ-point and a time step of 1.5 fs was used. We employed the generalized gradient approximation to the exchange-correlation functional due to Becke-Lee-Yang-Parr (BLYP) [53] that was shown to reproduce well the structural properties of liquid and amorphous GeSe$_2$ and GeSe compounds [54].

The amorphous GAS234 models contain 300 atoms in cubic box with an edge of 20.413 Å, corresponding to a density of 0.0352 atoms/Å$^3$ which was assigned in turn by interpolation of the experimental densities from neighbor compositions given in Refs. [55]. The amorphous models (M1-M4) were generated by quenching from 1200 K to 300 K in about 120 ps. The atomic positions of the a-GAS234 models equilibrated at 300 K were then optimized at zero temperature. The model M1 and M4 are actually not fully independent. In fact, they were generated in the last step of quenching from 500 K to 300 K from two snapshots of the same trajectory at 500 K separated by a short time interval. We followed this procedure to check the dependence of the structural and electronic properties of the models on the very last step of the quenching. Models M1 and M4 are structurally rather similar, and both feature the same structural unit consisting of a As-Se-As-As-As-Ge chain on which the LUMO state is localized. However, the second coordination shell of the atoms belonging to this chain differs in the two models which leads to a different response of the LUMO state to electron charging and to the electric field as we discussed in the article.

To quantify the localization properties of individual KS states, we computed the inverse participation ratio (IPR) which is defined for the i-th KS state by $\sum_j c_{ij}^4 / \left(\sum_j c_{ij}^2\right)^2$, where j runs over the GTO of the basis set and $c_{ij}$ are the expansion coefficients of the i-th KS state in GTOs. The IPR takes values varying from 1/N for a completely delocalized electron, where N is the number of atomic-like orbitals in the basis set of the whole supercell, to one for an electron completely localized on a single atomic-like orbital.

A uniform electric field is introduced in the total energy as a term -**P·F** where **P** is the dipole moment computed within the Berry's phase approach [43]. The dipole moment is defined only modulo the quantum of polarization which amounts to transfer an electron by the primitive Bravais vectors of the supercell. This scheme allows introducing a uniform electric field by still keeping the periodic boundary conditions. This scheme can be applied in the presence of an electronic gap between delocalized states sufficiently large to prevent the Zener breakdown.

## References

[1] S. R. Ovshinsky, Phys. Rev. Lett. 21, 1450–1453 (1968).

[2] A. Pirovano, A. Lacaita, A. Benvenuti, F. Pellizzer, R. Bez, IEEE Trans. Electron Devices 51, 452–459 (2004); M. Wuttig, N. Yamada, Nature Mater. 6, 824 (2007); P. Noé, C. Vallée, F. Hippert, F. Fillot, J.-Y. Raty, Semicond. Sci. Technol. 33, 013002 (2018); W. Zhang, R. Mazzarello, M. Wuttig, E. Ma, Nat. Rev. Mater. 4, 150 (2019).

[3] P. Fantini, J. Phys. D: Appl. Phys. 53, 283002 (2020).

[4] S.W. Fong, C. M. Neumann, H.-S. P. Wong, IEEE Trans. Electron Devices 64, 4374 (2017).

[5] P. Cappelletti, R. Annunziata, F. Arnaud, F. Disegni, A. Maurelli, P. Zuliani, J. Phys. D: Appl. Phys. 53, 193002 (2020); A. Redaelli, E. Petroni, R. Annunziata, Mater. Sci. Semicond. Process. 137, 106184, (2022).

[6] D. Kuzum, R. G Jeyasingh, B. Lee, and H.-S. P. Wong, Nano Lett. 12, 2179–2186 (2012); T. Tuma, A. Pantazi, M. Le Gallo, A. Sebastian and E. Eleftheriou, Nat. Nanotechnol. 11, 693–699 (2016); A. Sebastian, M. Le Gallo, R. Khaddam-Aljameh, and E. Eleftheriou, Nat. Nanotechnol. 15, 529–544 (2020).

[7] H.-Y. Cheng, F. Carta, W. C. Chien, H.-L. Lung, M. J. BrightSky, J. Phys. D: Appl. Phys. 52, 473002 (2019).

[8] Z. Zhao, S. Clima, D. Garbin, R. Degraeve, G. Pourtois, Z. Song, M. Zhu, Nano-Micro Letters 16, 81 (2024).

[9] D. Kau, S. Tang, I. V. Karpov, R. Dodge, B. Klehn, J. A. Kalb, J. Strand, A. Diaz, N. Leung, J. Wu, S. Lee, T. Langtry, K. Chang, C. Papagianni, J. Lee, J. Hirst, S. Erra, E. Flores, N. Righos, H. Castro, G. Spadini, 2009 IEEE International Electron Devices Meeting (IEDM), pp. 1-4.

[10] A. Fazio, 2020 IEEE International Electron Devices Meeting (IEDM), pp. 24.1.1-24.1.4.

[11] M. Alayan, et al., 2017 International Electron Devices Meeting (IEDM), pp. 2.3.1–2.3.4.

[12] I. Tortorelli, S. Tang, and C. Papagianni, US10134470B2, Nov. 20, 2018.

[13] T. Ravsher, R. Degraeve, D. Garbin, A. Fantini, S. Clima, G. L. Donadio, S. Kundu, H. Hody, W. Devulder, J. Van Houdt, V. Afanas'ev, R. Delhougne, G. S. Kar, 2021 IEEE International Electron Devices Meeting (IEDM), pp. 28.4.1–28.4.4.

[14] T. Ravsher et al., Phys. Status Solidi RRL 17, 2200417 (2023).

[15] S. Hong, H. Choi, J. Park, Y. Bae, K. Kim, W. Lee, S. Lee, H. Lee, S. Cho, J. Ahn, S. Kim, T. Kim, M.-H. Na, S. Cha, 2022 IEEE International Electron Devices Meeting (IEDM), pp. 18.6.1–18.6.4.

[16] F. Pellizzer et al., 2023 IEEE International Electron Devices Meeting (IEDM), pp. 1-4; doi: 10.1109/IEDM45741.2023.10413669

[17] I.W. Park et al 2023 IEEE International Electron Devices Meeting (IEDM), pp. 1-4, doi: 10.1109/IEDM45741.2023.10413748

[18] T. Ravsher, D. Garbin, A. Fantini, R. Degraeve, S. Clima, G. L. Donadio, S. Kundu, H. Hody, W. Devulder, G. Potoms, T. Peissker, L. Nyns, J. Van Houdt, V. Afanas'ev, A. Belmonte, and G. Sankar Kar, Phys. Status Solidi RRL 18, 2300415 (2024).

[19] T. Ravsher, R. Degraeve, D. Garbin, S. Clima, A. Fantini, G. Donadio, S. Kundu, W. Devulder, H. Hody, G. Potoms, J. Van Houdt, V. Afanas'ev, A. Belmonte, G. Kar, 2024 IEEE International Reliability Physics Symposium (IRPS), 7A. 5-1-7A. 5-9, 2024.

[20] S. Clima, F.Ducry, D. Garbin, T. Ravsher, R.Degraeve, A. Belmonte, G. Sankar Kar, and G. Pourtois, 2024 IEEE International Electron Devices Meeting (IEDM), pp. 1-4,

[21] S. Clima, D. Garbin, K. Opsomer, N.S. Avasarala, W. Devulder et al., Phys. Status Solidi RRL 14, 1900672 (2020).

[22] S. Clima, T. Ravsher, D. Garbin, R. Degraeve, A. Fantini, R. Delhougne, G. Sankar Kar, and G. Pourtois, ACS Appl. Electron. Mater. 5, 461–469 (2023).

[23] J. Lee, Y. Seo, S. Ban, D. G. Kim, Y. B. Park, T. H. Lee, and H. Hwang, IEEE Trans. Electron Devices 71, 3351 (2024).

[24] H. J. Sung, M. Choi, Z. Wu, H. Chae, S. Heo, Y. Kang, B. Koo, J. B. Park, W. Yang, Y. Park, Adv. Sci. 2024, 11, 2408028.

[25] P. Fantini et al., Adv. Electr. Mat. 9, 2300037 (2023).

[26] P. Fantini, A. Ghetti, E. Varesi, A. Pirovano, D. Baratella, C. Ribaldone, D. Campi, M. Bernasconi and R. Bez, 2024 IEEE International Electron Devices Meeting (IEDM), pp. 1-4.

[27] P. Fantini, M. Ferro, and A. Calderoni, Appl. Phys. Lett. 102, 253505 (2013).

[28] D.S. Jeong et al., J. App. Phys. 98, 113701 (2005).

[29] V. M. Arutyuntan et al. Infrared Phys. 29, 681 (1988).

[30] D. Lencer, M. Salinga, B. Grabowski, et al. et al. Nature Mater 7, 972–977 (2008).

[31] H.Y. Cheng, W.C. Chien, I.T. Kuo, C.W. Yeh, L. Gignac et al., 2018 IEEE International Electron Devices Meeting (IEDM), pp. 37.3.1– 37.3.4; H.Y. Cheng, W.C. Chien, I.T. Kuo, C.H. Yang, Y.C. Chou et al., 2021 IEEE International Electron Devices Meeting (IEDM), pp. 28.6.1– 28.6.4.

[32] N. Marzari and D. Vanderbilt, Phys. Rev. B 56, 12847 (1997).

[33] D. D. Vaughn, R. J. Patel, M. A. Hickner, R. E. Schaak, J. Am. Chem. Soc. 132, 15170 (2010).

[34] G. Dittmar and H. Schaefer, Acta Crystallogr. B 32, 2726 (1976).

[35] K. Shimakawa, J. Non-Cryst. Solids 43, 229 (1981).

[36] C. Jacoboni, E. Piccini, R. Brunetti, and M. Rudan, J. Phys. D 50, 255103 (2017).

[37] X. Zhang, K. Li, J. Zhou, S. R. Elliott, and Z. Sun, Adv. Electron. Mater. 2400291 (2024).

[38] A. Slassi, L.-S. Medondjio, A. Padovani, F. Tavanti, X. He, Sergiu Clima, Daniele Garbin, Ben Kaczer, Luca Larcher, Pablo Ordejón, and Arrigo Calzolari, Adv. Electron. Mater. 9, 2201224 (2023).

[39] M. Xu, M. Xu, X. Miao, InfoMat 4, e12315 (2022).

[40] J.-Y. Raty, P. Noé, Phys. Status Solidi RRL 14, 2070024 (2020).

[41] P. Noé, A. Verdy, F. d'Acapito, J.-B. Dory, M. Bernard, G. Navarro, J.-B. Jager, J. Gaudin, J.-Y. Raty, Sci. Adv. 6, eaay2830 (2020).

[42] Y. Guo, H. Li, W. Zhang, and J. Robertson, Appl. Phys. Lett. 115, 163503 (2019).

[43] I. Souza, J. Iñiguez, and D. Vanderbilt, Phys. Rev. Lett. 89, 117602 (2002); P. Umari and A. Pasquarello, Phys. Rev. Lett. 89, 157602 (2002).

[44] Marius Grundmann, The Physics of Semiconductors, Springer International Publishing Switzerland, 2016.

[45] M. Nardone, M. Simon, I. V. Karpov, V. G. Karpov, J. Appl. Phys. 112, 071101 (2012).

[46] K. Konstantinou, F.C. Mocanu, J. Akola, S.R. Elliott, Acta Mater. **223**, 117465 (2022).

[47] Sentaurus TCAD tools, Synopsys, Mountain View, CA USA 2022, https://www.synopsys.com/manufacturing/tcad/device-simulation/sentaurus-device.html

[48] J. VandeVodele, M. Krack, F. Mohamed, M. Parrinello, T. Chassaing, J. Hutter, Comput. Phys. Commun. 167, 103 (2005).

[49] S. Goedecker, M. Teter, J. Hutter, Phys. Rev. B 54, 1703 (1996).

[50] T. D. Kühne, M. Krack, F. R. Mohamed, and M. Parrinello, Phys. Rev. Lett. 98, 066401 (2007); T. D. Kühne, M. Krack, and M. Parrinello, J. Chem. Theory Comput. 5, 235 (2009).

[51] R. Car and M. Parrinello, Phys. Rev. Lett. 55, 2471 (1985).

[52] S. Caravati, M. Bernasconi, T. D. Kühne, M. Krack, and M. Parrinello, Appl. Phys. Lett. 91, 171906 (2007); S. Caravati, M. Bernasconi, and M. Parrinello, Phys. Rev. B 81, 014201 (2010); E. Spreafico, S. Caravati, and M. Bernasconi, Phys. Rev. B 83, 144205 (2011).

[53] A. D. Becke, Phys. Rev. A 38, 3098 (1988); C. Lee, W. Yang, and R. G. Parr, Phys. Rev. B 37, 785 (1988).

[54] M. Micoulaut, R. Vuilleumier, and C. Massobrio, Phys. Rev. B 79, 214205 (2009); S. Le Roux, A. Bouzid, M. Boero, and C. Massobrio, J. Chem. Phys. 138, 174505 (2013).

[55] R. P. Wang, A. Smith, B. Luther-Davies, H. Kokkonen, and I. Jackson, J. Appl. Phys. 105, 056109 (2009); T. Wang, W. H. Wei, X. Shen, R. P. Wang, B. Luther Davies, and I. Jackson, J. Phys. D: Appl. Phys. 46, 165302 (2013).

# SUPPLEMENTARY INFORMATION

| alloy | Se (atom %) | Band gap (eV) |
|---|---|---|
| ch-rich | 64 | 1.92 |
| ch-poor | 54 | 1.8 |

**Table S1.** Bilayer stack: chalcogen composition content (at. %) and corresponding optical bandgap.

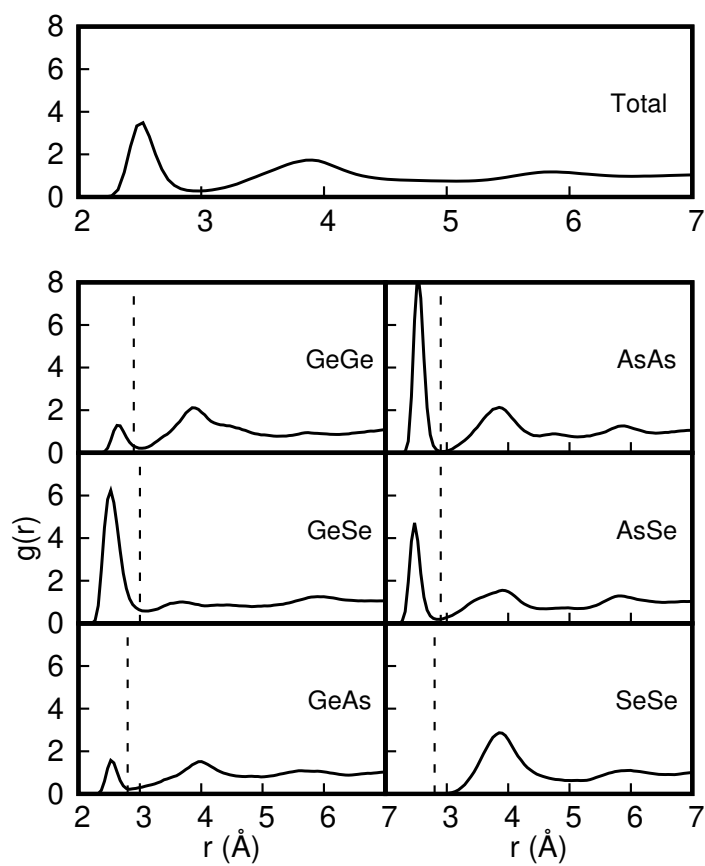

**Figure S1.** Partial and total pair correlation functions of amorphous GAS234 (a-GAS234) at 300 K averaged over four independent 300-atom models. The vertical lines indicate the cutoff used to compute the coordination numbers.

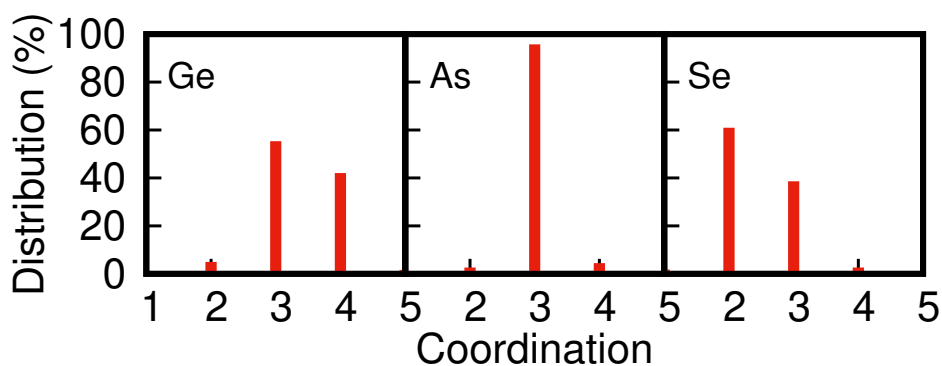

**Figure S2**. Distribution of coordination numbers for the different species, averaged over four a-GAS234 models.

|    | Ge   | As   | Se   | Total |
|----|------|------|------|-------|
| Ge | 0.28 | 0.32 | 2.78 | 3.38  |
| As | 0.27 | 1.46 | 1.29 | 3.02  |
| Se | 1.54 | 0.86 | 0.00 | 2.40  |

**Table S2**. Distribution of the partial coordination numbers averaged over four a-GAS234 models.

|   | AsSe | GeSe | AsAs | GeAs | GeSe | GeGe |
|---|------|------|------|------|------|------|
| % | 27   | 49   | 15.5 | 5.7  | 0.0  | 2.5  |

**Table S3.** Percentage of the different types of bonds averaged over four a-GAS234 models.

|   | $Ge_t$ | isolated $Ge_t$ | Corner-sharing $Ge_t$ | Edge-sharing $Ge_t$ |
|---|--------|-----------------|-----------------------|---------------------|
| % | 33     | 60              | 35                    | 4                   |

**Table S4.** Percentage of Ge atoms in a tetrahedral environment ($Ge_t$) with respect to the total number of Ge atoms, and percentage of isolated, corner- and edge-sharing tetrahedra. Isolated means that the tetrahedron does not share vertices with other tetrahedra. The data are averaged over four independent models. The Ge atoms in a tetrahedral environment was computed from the order parameter for tetrahedricity introduced in Ref. [S1] and used in our previous works on amorphous tellurides [S2].

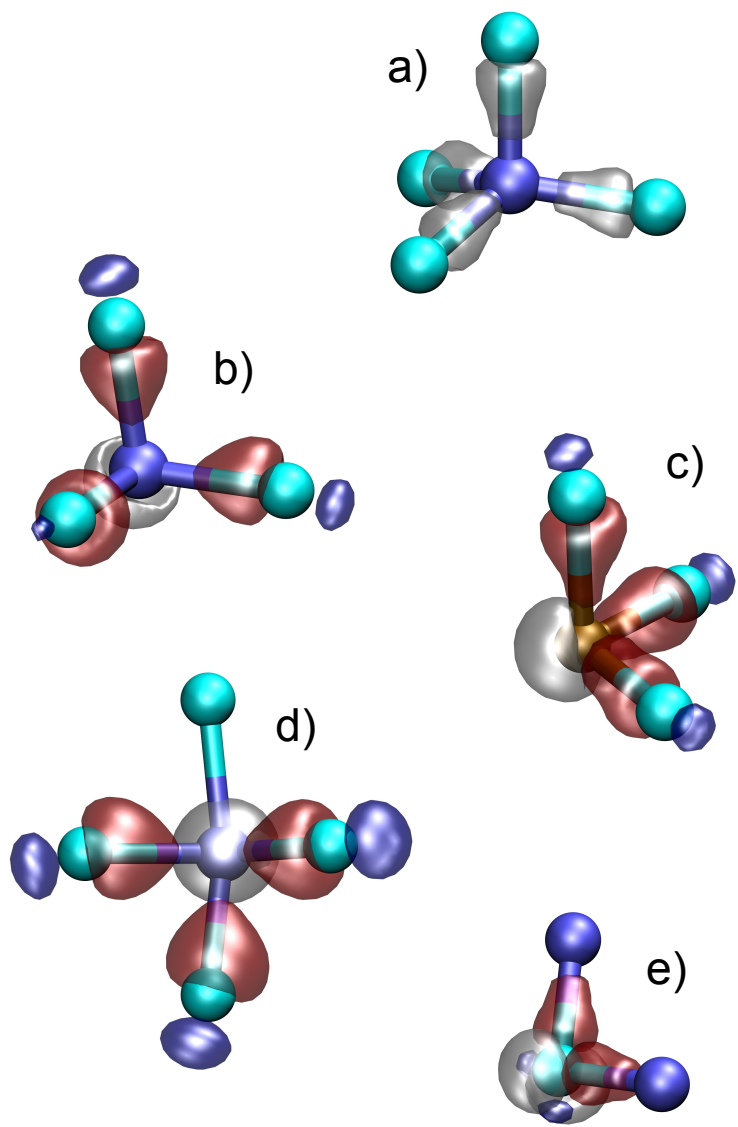

**Figure S3.** Isosurface of Wannier functions (WF) for different atomic environments in the models of a-GAS234. WFs are the periodic version of the occupied Boys orbitals obtained by the unitary transformation of the Kohn-Sham occupied orbitals, which minimizes the quadratic spread [S3]. a) Ge in a tetrahedral site, b) Ge and c) As in a pyramidal bonding geometry, d) Ge in defective octahedra with coordination four and e) Se two-coordinated. The pyramidal geometry of three-coordinated Se is the same of that of panel c). The atomic color code is the same as in Figure 6a in the article. Iso-surfaces with different colors (red and blue) have different sign. Wannier functions with spherical iso-surfaces (gray) are s-type lone pairs.

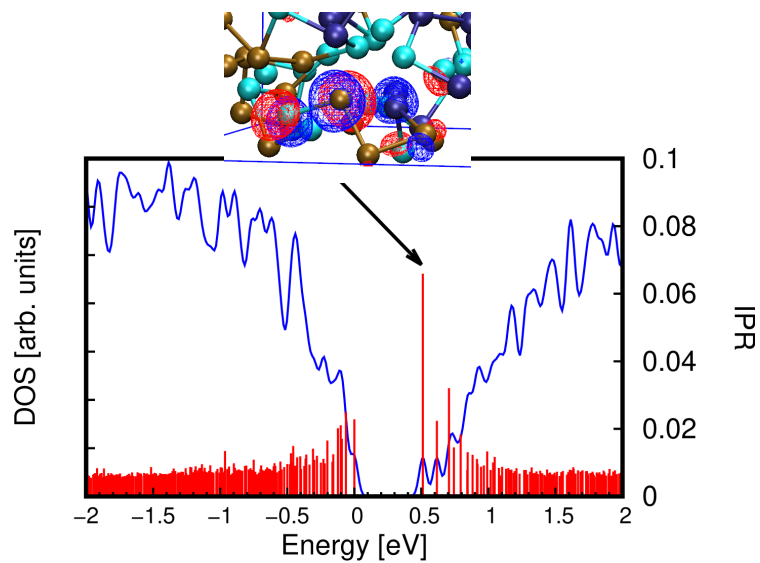

**Figure S4.** Electronic density of states (DOS) and inverse participation ratio (IPR) of model M1 of a-GST234. The zero of energy corresponds to the highest occupied state (top of the valence band or highest occupied molecular orbital, HOMO). The DOS is computed from KS energies at the supercell Γ-point with a 27 meV Gaussian broadening. The iso-surface of an in-gap localized electronic state (lowest unoccupied molecular orbital, LUMO) is highlighted in the inset. Iso-surfaces with different colors (red and blue) have different sign. The LUMO is localized over a chain of wrong As-As and As-Ge bonds (As-Se-As-As-As-Ge chain, see Figure 6a for the color code on atoms).

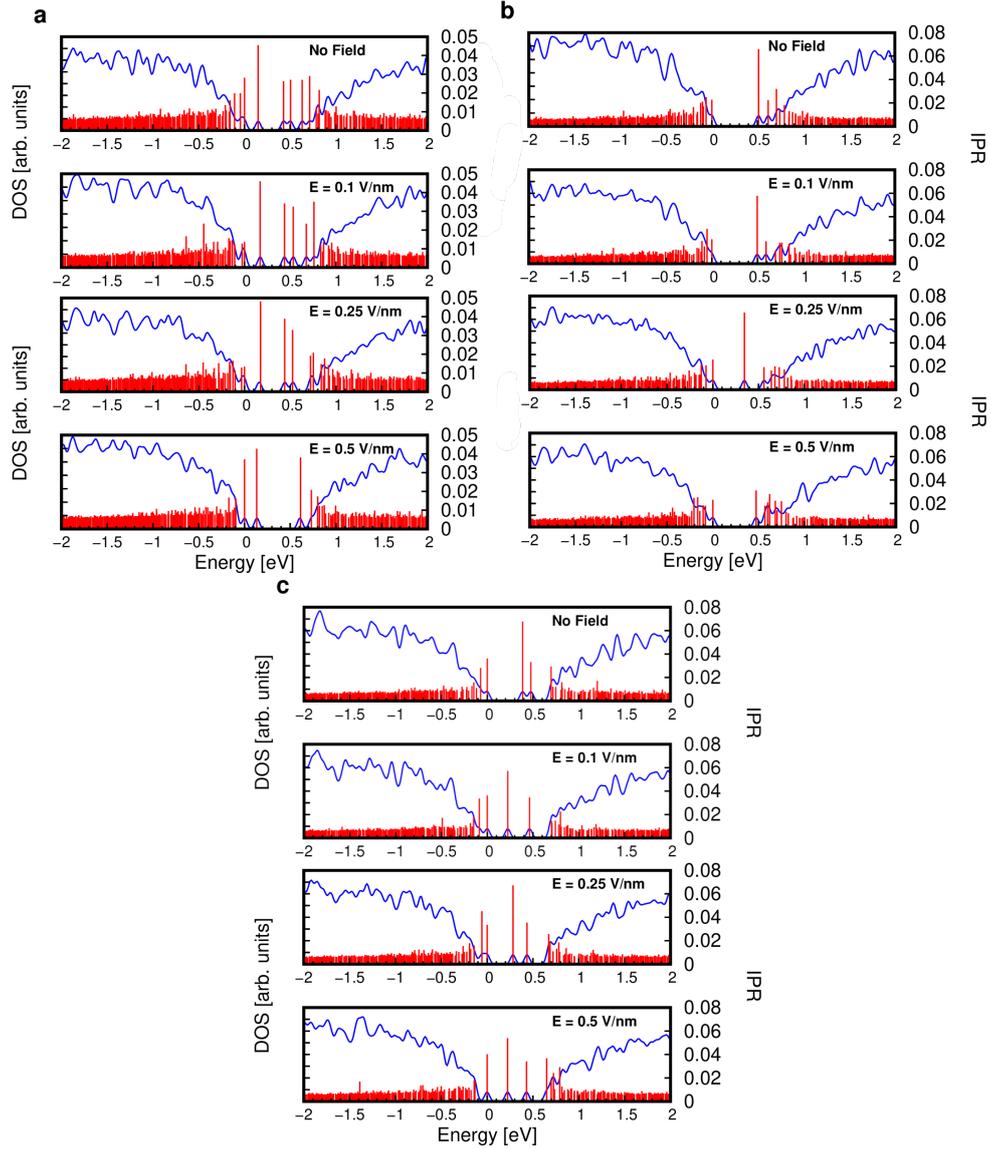

**Figure S5.** Electronic density of states (DOS) and inverse participation ratio (IPR) of model a) M3 b) M1 c) M2 of a-GST234. The models are relaxed with no field or with an electric field and then further relaxed by removing the field. The strength of the field is given in the insets. The zero of energy always corresponds to the highest occupied Kohn-Sham state (highest occupied molecular orbital, HOMO). In model M3 (panel a) no sizable changes are observed with a field up to 0.25 V/nm. At 0.5 V/nm we observed instead a dramatic change of the states in the gap, the lowest unoccupied molecular orbital (LUMO) is now close to the valence band and the LUMO+1 is close to the conduction band. However, these two states are not related to the LUMO and LUMO+1 states of the original model. The LUMO is now localized on an As-As-As-Ge chain while the LUMO+1 is now localized on a GeSe$_4$ group with a tetrahedral Ge atom. Therefore, a large electric field leads to the removal of deep defect states in the gap in model M3. In model M1 (panel b), one observes a slight shift to lower energy of the LUMO state for a field strength up to 0.25 V/nm. This state is localized on an As-Se-As-As-As-Ge chain (see article). At the highest field of 0.5 V/nm, the LUMO and LUMO+1 states which were originally localized of neighboring chains of wrong bonds mix to some extent. The LUMO moves to a position in energy closer to the original one, but a few other localized states at the edge of the conduction band move to slightly lower energy. In model M2 (panel c), the LUMO state does not change sizably with the field up to 0.25 V/nm. At 0.5 V/nm, the LUMO shifts close to the valence band by keeping the same character (it is localized on a Se$_3$Ge-Ge group). The HOMO state changes instead its character, and it lifts off the valence edge.

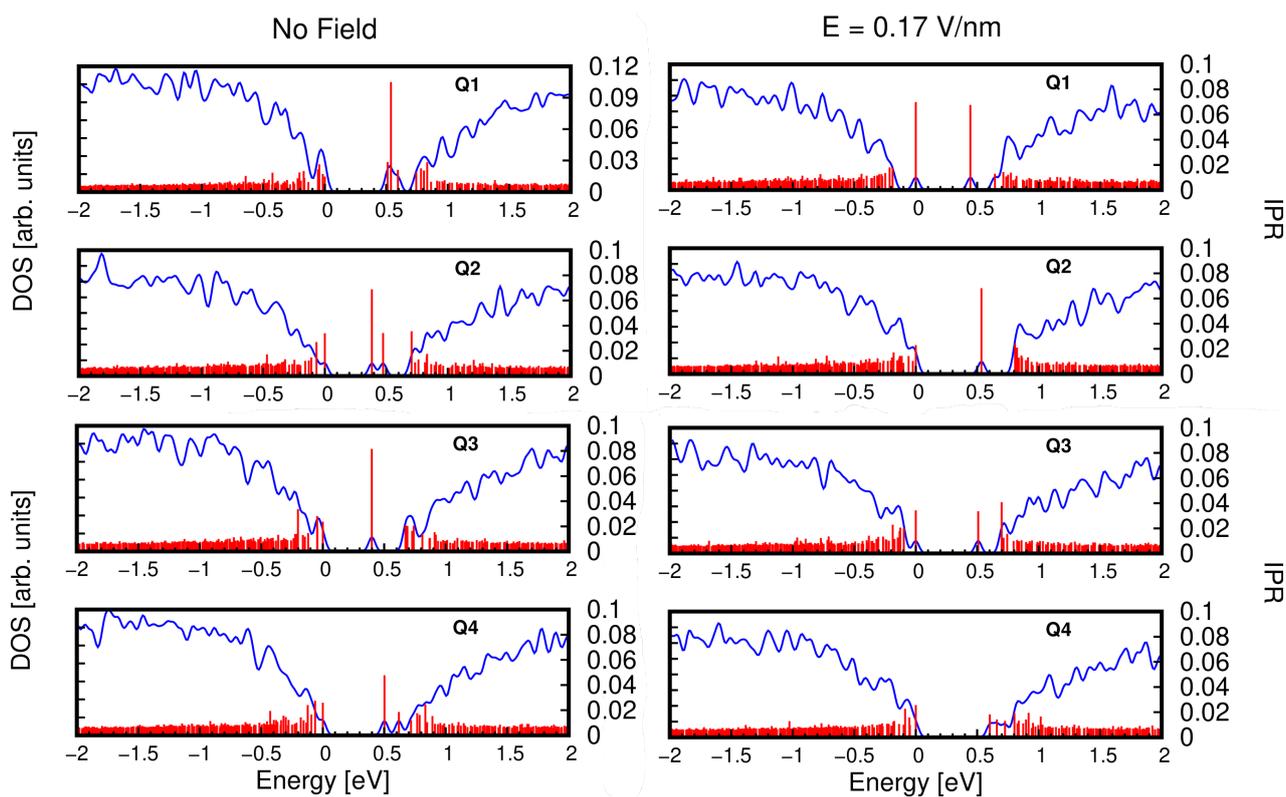

**Figure S6.** Electronic density of states superimposed to the IPR for the four models Q1-Q4 equilibrated at 500 K with (right panels) and without (left panels) the electric field 0.17 V/nm large. All the models were then quenched at 300 K and finally relaxed at zero temperature without the electric field and the DOS are computed without the electric field as well. The zero of energy always corresponds to the highest occupied Kohn-Sham state (highest occupied molecular orbital, HOMO).

## References

[S1] J. R. Errington and P. G. Debenedetti, Nature 409, 318 (2001)

[S2] E. Spreafico, S. Caravati, and M. Bernasconi, Phys. Rev. B 84, 144205 (2011).

[S3] N. Marzari and D. Vanderbilt, Phys. Rev. B 56, 12847 (1997)